\documentclass[11pt]{article}
\usepackage{moriond,epsfig,graphicx}

\def\be{\begin{equation}}
\def\ee{\end{equation}}
\def\bea{\begin{eqnarray}}
\def\eea{\end{eqnarray}}

\begin{document}
%%%%%%%% HAS BEEN SUPPRESSED \vspace*{4cm}
\title{Universal Scrambling Properties of Spectra \\
and Wave functions in Disordered Interacting Systems}

\author{\underline{Fr\'ed\'eric Pi\'echon},
Gilles Montambaux}

\address{Laboratoire de Physique des Solides B\^at 510, CNRS, 
Universit\'e Paris-Sud, 91405 Orsay, France}

\maketitle\abstracts{
Recent experiments on quantum dots in the Coulomb Blockade regime 
have shown how adding
successive electrons into a dot modifies the energy spectrum and the wave
functions of the electrons already present in the dot. 
Using a microscopic model, we study the
importance of electron-electron interaction on these ``scrambling'' effects.
We compute the Hartree-Fock single
particle properties as function of the number $p$ of added electrons. 
We define parametric correlation functions that characterize
the scrambling properties of the Hartree-Fock wave functions and energy
spectra. We find that each of these correlation functions exhibit a
universal behavior in terms of the ratio $p/p^{\ast }$
where $p^{\ast }$ is a characteristic number that
decreases with increasing either the interaction strength, the disorder
strength or the system size.}

\section{Introduction}

At very low temperature, by applying a gate voltage $V_g$ 
to a Quantum Dot (QD) weakly connected to leads by tunnel contacts, 
one obtains a series of Coulomb blockade conductance peaks of height $G_{N}$
at positions $V_{g}^{N}$. The peak positions $V_{g}^{N}$ are directly 
related to difference 
$E_{N+1}-E_{N}$ between the ground state energies of the dot with $N$ and 
$N+1$ electrons (in zero bias). The heights $G_{N}$ are related to
the overlap of the $|N>$ and $|N+1>$ 
ground state wave functions at the contacts \cite{review}.
During the last few years, many experiments \cite{additionexptheo,additionexp} 
have shown that the statistical properties
of the successive peak positions $V_{g}^{N}$ could not
be described within the framework of the random matrix theory (RMT). 
%for the (non-interacting) chaotic single particle motion.
For the addition spectrum 
(e.g. statistical properties of the differences 
$\Delta _{2}=V_{g}^{N+1}-V_{g}^{N}$), qualitative agreement
with experimental results can only be obtained 
by taking into account fluctuations of
the charging energy that originate from the spatial fluctuations of wave
function amplitudes \cite{additionexptheo,additionRMT,additionSCHF}.
Although such fluctuations of wave functions already exist in RMT \cite
{additionRMT}, the resulting charging energy
fluctuations are parametrically to small if one ignores the reorganization 
or ``scrambling'' (of the wave functions and of the spectrum) induced by the
addition of the successive electrons \cite{additionexptheo,additionSCHF}. 
In fact, in a recent experimental study of the peak to peak heights
correlation, Patel et {\em al.} \cite{scramblingexp} 
have unambiguously shown that
this reorganization is important and could essentially 
be attributed to electron-electron interaction effects. 
Recently, Alhassid et {\em al.} \cite{scramblingtheo} have confirmed 
the importance of scrambling using a phenomenological model 
based on Gaussian Random Matrix Process (GRMP)\cite{grmp},
and have obtained good agreement 
with Patel et {\em al.} experiment.

In the present work based on a microscopic model of a
quantum dot, we present a detailed numerical study that shows directly the
importance of the electron-electron interaction on the scrambling of the
single particle properties obtained within a self consistent Hartree-Fock
(SCHF) method. 

\section{Microscopic model}
We model a two dimensional
disordered dot of size $A=L_{x}\times L_{y}$ by the following
Hamiltonian: 
\begin{equation}
H=\sum_{i}w_{i}c_{i}^{\dag }c_{i}-t\sum_{<ij>}[c_{i}^{\dag }c_{j}+h.c.]+
\frac{U}{2}\sum_{i,j=1}^{A}M_{ij}n_{i}n_{j},  \label{hamiltonian}
\end{equation}
where $<ij>$ denotes nearest neighbors on a square lattice of size 
$A=L_{x}\times L_{y}$ and $t=1$. $c_{i}^{\dag }$ and $c_{i}$ are creation and
annihilation operators for an electron (spinless) on site $i$ and 
$n_{i}=c_{i}^{\dag }c_{i}$. The random on-site disorder potential $w_{i}$ are
uniformly distributed over $[-\frac{W}{2},\frac{W}{2}]$. Because of the
disorder, the electronic charge density exhibits fluctuations in space. 
We consider that the natural bare 
interaction between these charge fluctuations has
a long range Coulomb form $M_{ij}=\frac{1}{|{\bf r}_{i}-{\bf r}_{j}|}$. We
choose a torus geometry to prevent from electrostatic 
effects that tend to accumulate the charge
on the surface. The distance $|{\bf r}_{i}-{\bf r}_{j}|$ is calculated 
from the shortest path relating the two sites $i$ and $j$ on the torus.

\section{Hartree Fock picture of scrambling}

We treat interaction effects in the self-consistent
Hartree-Fock approximation. We thus assume that the ground
state $|\Psi _{0}^{N}>$ of a dot with $N$ particles is a Slater determinant of 
$N$ effective single particle states $\psi _{\alpha }^{N}({\bf r}_{i})$ with
associated quasi-energy $\epsilon _{\alpha }^{N}$. 
These states $\psi _{\alpha }^{N}({\bf r}_{i})$ 
are in fact the self-consistent eigenstates 
of an effective one--particle Hamiltonian 
$h^{N}=\sum_{i}w_{i}^{eff}c_{i}^{\dag
}c_{i}-\sum_{<ij>}[t_{ij}^{eff}c_{i}^{\dag }c_{j}+h.c.]$. The
effective onsite potential $w_{i}^{eff}=w_{i}+U\sum_{j}M_{ij}<\Psi _{0}^{N}|n_{j}|\Psi _{0}^{N}>$ is
the sum of the disorder potential and the Hartree contribution. Similarly the effective hopping $t_{ij}^{eff}=t\delta _{<ij>}+UM_{ij}<\Psi _{0}^{N}|c_{i}^{\dag}c_{j}|\Psi _{0}^{N}>$ includes the Fock exchange term. 
Self consistency is obtained with an iterative procedure.

The self consistent states $\psi_{\alpha}^{N}({\bf r}_{i})$
are linear combinations of the non interacting 
$\psi_{\alpha}^{0}({\bf r}_{i})$. However the $\psi_{\alpha}^{N}({\bf r}_{i})$ 
depend explicitly on interaction strength and particle number $N$ 
and not only on disorder strength and
system size as the $\psi _{\alpha }^{0}({\bf r}_{i})$. 
Therefore, by adding one by one $p$ electrons 
into a dot with initially $N$ particles, the
effective one body Hamiltonian $h^{N}$ is changed into another Hamiltonian 
$h^{N+p}$ with the
corresponding $\psi _{\alpha}^{N+p}({\bf r}_{i})$ and 
$\epsilon _{\alpha }^{N+p}$. In
the Hartree-Fock approximation, one can always write $h^{N+p}=h^{N}+V^{N,N+p}$
, such that the scrambling effects of the $p$ additional electrons is described
by an effective one body term $V^{N,N+p}$, that is neither diagonal in the
position basis nor in the $\psi _{\alpha }^{N}$ basis 
of the dot with $N$ particles. 
In fact this possibility of writing $h^{N+p}=h^{N}+V^{N,N+p}$ allows 
us to make a parallel with the phenomenological model of Alhassid 
and Malhotra \cite{scramblingtheo}. In their work, the $N+p$ interacting 
particles system is described by an
effective one body Hamiltonian 
$h^{N+p}=h_{0}\cos{(p\delta x)}+h_{1}\sin{(p\delta x)}$ 
in which $h_{0},h_{1}$ are two random Gaussian matrices and $\delta x$
is the parameter that control the scrambling strength. 
Their effective scrambling potential $V^{N,N+p}$ is thus given by 
$V^{N,N+p}\simeq h_0(\cos{(p\delta x)}-1)+h_{1}\sin(p\delta x)$. 

\section{Numerical results}
\subsection{Scrambling properties of SCHF single particle wave functions}
In order to characterize the scrambling of the SCHF single particle wave functions, we have computed the following parametric correlation function:
\begin{equation}
C_{\psi}(p)= \overline{\left|\int d{\bf r\ }\psi_{\alpha }^{N+p}({\bf r}
)\psi _{\alpha }^{N\ast}({\bf r})\right|^2}
\end{equation}
in which $\overline{\ldots }$ means averaging on disorder configurations 
and on the $N$ lowest single particle energy levels $\alpha$. 
The scrambling of wavefunctions is characterized 
by a decrease of $C_{\psi }(p)$ with increasing $p$. 
A similar quantity 
has already been studied in the context of GRMP \cite{grmp}. 
We have studied in detail the dependences of $C_{\psi }(p)$ on 
the disorder strength $W$ (or conductance $g$), the interaction strength $U$
and the system size $A$. The ranges of parameters considered are $W=4 \rightarrow 10$; $U=0.5 \rightarrow 3$, $A=6 \times 6 \rightarrow 11\times 10$ sites. The initial values of $N$ are chosen so that the band is quarter filled $\nu=N/A \simeq 1/4$. For each set of parameters we have averaged over 500 distinct disorder configurations. The results obtained for $C_{\psi }(p)$ are summarized on Figure 1a. This figure shows a quite interesting result, that the correlation function $C_{\psi}(p)$ can be recast into a one parameter scaling form $C_{\psi}(p)=f_{\psi}(p/p_{\psi})$ where $p_{\psi}$ depends on the three parameters 
$W,U,A$. In the range of parameters studied we further obtain the following
approximate fitting form:
\begin{equation}
\begin{array}{ll}
C_{\psi }(p)&\simeq
{\frac{\displaystyle{1}}{\displaystyle{[1+(\frac{p}{p_{\psi }})^{3/4}]^{1/2}}}}\\
1/p_{\psi}&= c \ U^2 A^2 W^{3} [1+ d_A \ U^2].
\end{array}
\end{equation} 
where $c\simeq 8. 10^{-8}$ and $d_A=0.03; 0.05; 0.1; 0.15; 0.2$ 
for $A=6 \times 6; 6 \times 7; 7 \times 8; 8\times 9; 9\times 10$ 
respectively. For a given $p$, $C_{\psi }(p)$ decreases 
when either $W,U$ or $A$ is increased.
Thus the larger the system size, the stronger is the scrambling
effect of a single additional electron. 
Let us compare the scaling $(3)$ with the scaling obtained by Alhassid and Malhotra in the context of the GRMP model \cite{scramblingtheo}. The behavior found in this model for the correlation function is $C_{\psi }^{GRMP}(p)\simeq
[1+(\frac{p}{p_{\psi }})^{2}]^{-1}$ with $1/p_{\psi}\simeq \delta x \sqrt{M}$ for gaussian orthogonal matrices of size $M \times M$. 
%The size dependency of $1/p_{\psi}$ is stronger in our microscopic model ($\propto A^2$ instead of $\propto \sqrt{A}$). Moreover the overall size dependency of $C_{\psi }(p)$ (at large $p/p_{\psi}$) is weaker in the microscopic model ($\propto A^{-3/4}$ instead of $\propto A^{-1}$). 
%Since the function $C_{\psi }(p)$ is different for the two models, it is not possible to clearly identify an effective parameter $\delta x$ that would describe the microscopic parameters.
The difference between the functional form $C_{\psi }(p)$ and  $C_{\psi }^{GRMP}(p_{GRMP})$ seems to prevent any simple mapping between them. Nevertheless, if we consider only the tail of the two forms, we can identify the microscopic value $p$ with $p_{GRMP}^{16/3}$. We then obtain the following relation between the microscopic parameters and the parameters of the GRMP model : $(\delta x \sqrt{M})^{16/3} \sim c \ U^2A^2 W^3 (1+d_A \ U^2)$. 
\begin{figure}[h]
%\hspace{-1.5cm} \epsfig{figure=ww1.ps,width=9.cm}

%\hspace{-1.5cm} \epsfig{figure=wsca1.ps,width=9.cm}

\hspace{-1.cm} 
\includegraphics[scale=.52]{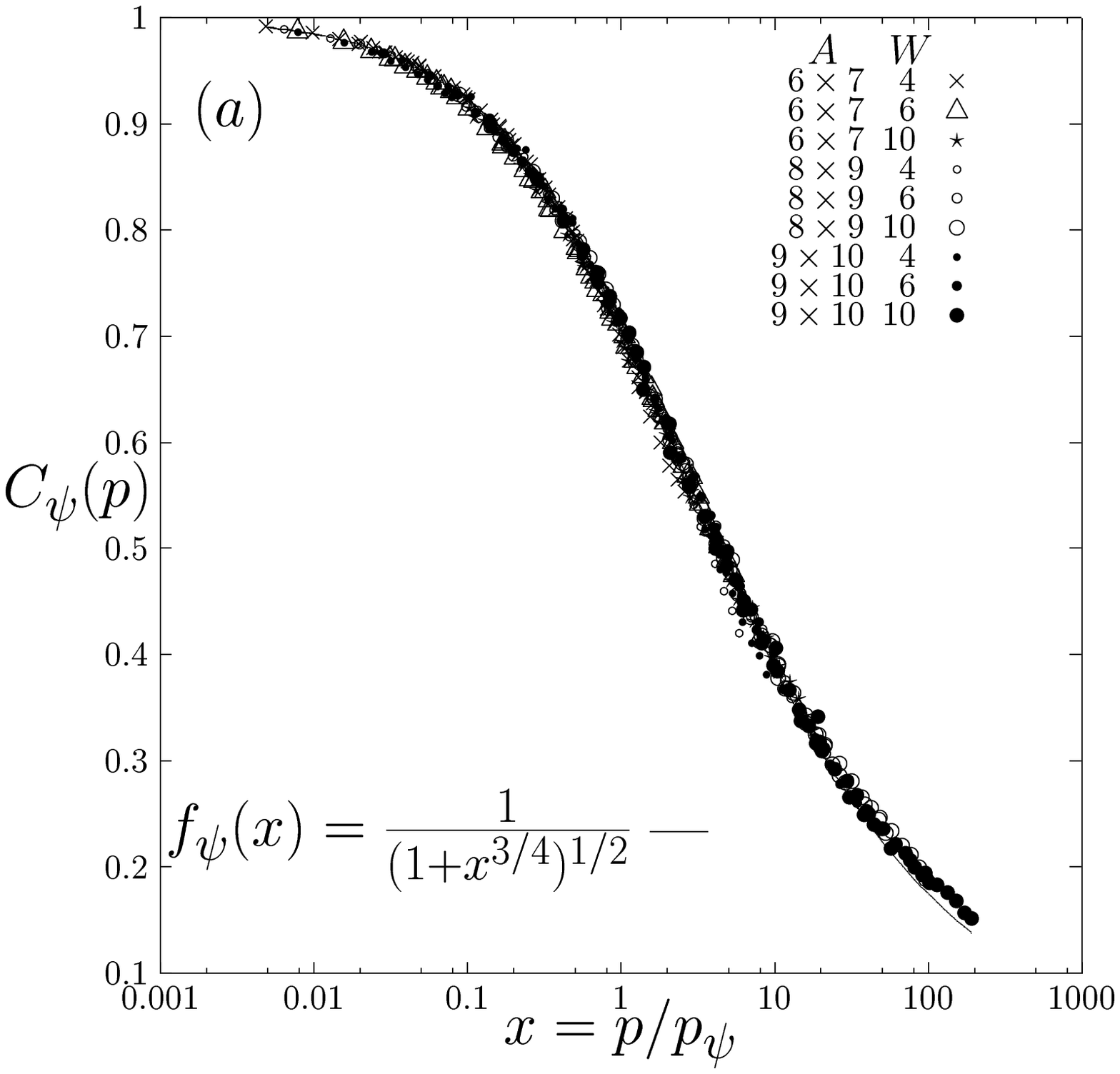}
\includegraphics[scale=.52]{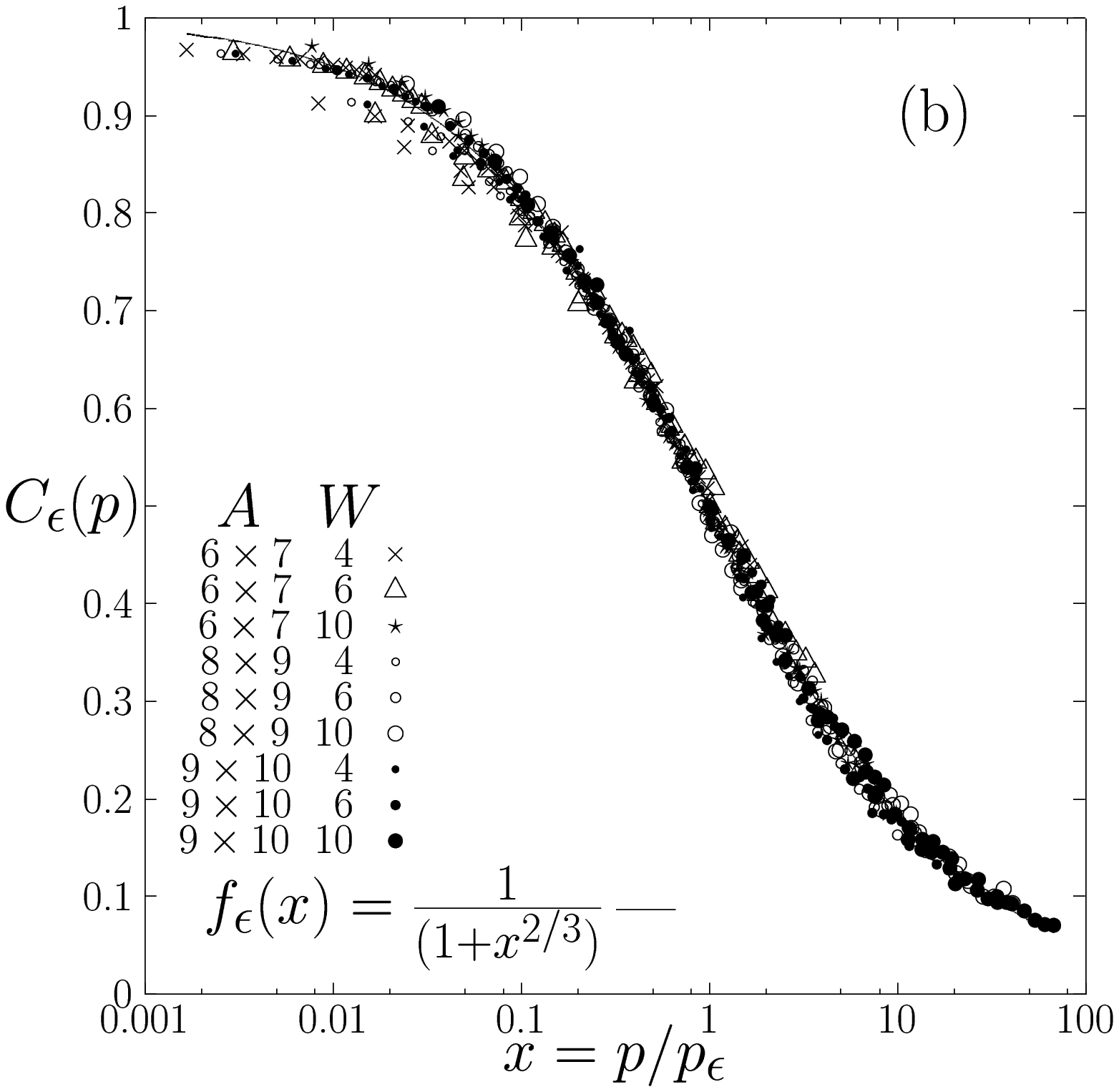}

%\vspace{-9.1cm} \hspace{+7.5cm} \epsfig{figure=ssca1.ps,width=9.cm}
\caption{
(a) $C_{\psi}(p)$ as a function of the scaling variable $p/p_{\psi}$. (b) $C_{\epsilon}(p)$ as a function of the scaling variable $p/p_{\epsilon}$. For a given disorder $W$ and system area $A$, we plot all the data points corresponding to the entire range $p=\lbrace 1,2,..,10 \rbrace$ and interaction range  $U=\lbrace 0.5,1,..,3\rbrace$.}
\end{figure}

\vspace{-0.75cm}
\subsection{Scrambling properties of SCHF single particle energy spectra}
In order to characterize the scrambling of the SCHF single particle 
energy spectrum we have computed the following parametric correlation function:
\begin{equation}
\begin{array}{l}
C_{\epsilon }(p)=
\frac{\displaystyle{\overline{s_{\alpha }^{N+p}s_{\alpha }^{N}}{\normalsize -1}}}{\displaystyle{\overline{s_{\alpha }^{N}s_{\alpha }^{N}}{\normalsize -1}}},
\end{array}
\end{equation}
where $s_{\alpha }^{N}=\epsilon _{\alpha +1}^{N}-\epsilon _{\alpha
}^{N}$ is the spacing between two consecutive levels 
($s_{\alpha }^{N}$ is normalized to the mean level spacing at this energy).  
Since the one particle density of state has a deep around the (average) Fermi level, we have used all $\alpha$ corresponding to energy levels 
(empty or occupied) neither in the this deep nor at the band edges. 
%If there is spectral scrambling, the nearest level
%spacings $s_{\alpha }^{N}$ vary with $p$ and thus $C_{\epsilon }(p)$
%changes with increasing $p$. 
As it is illustrated on Figure 1b, 
we have found that $C_{\epsilon }(p)$ can also be recast into a one parameter scaling form  $C_{\epsilon }(p)=f_{\epsilon }(p/p_{\epsilon})$ where $p_{\epsilon}$ depends on the parameters $W,U,A$. 
Quantitatively we obtain the following approximate fitting form:
\begin{equation}
\begin{array}{ll}
C_{\epsilon}(p)&\simeq 
\frac{\displaystyle{1}}{\displaystyle{1+(\frac{p}{p_{\epsilon }})^{2/3}}},\\
1/p_{\epsilon }&\simeq c U^{3} A^2 W^{2.75},
\end{array}
\end{equation}
where $c \simeq 5. 10^{-8}$. 
%The analytical form $1/p_{\epsilon }$ describes 
%the numerical data $1/p_{\epsilon }$ with less accuracy than 
%the fitting form $1/p_{\psi }$ describes the numerical data $1/p_{\psi}$.
The quantitative behavior of $p_{\epsilon}$, in terms of $W,A,U$, is different from that of $p_{\psi}$. This contrasts with GRMP-like models in which necessarily $p_{\epsilon}=p_{\psi}$. For the microscopic model, we believe that the difference found between $p_{\epsilon}$ and $p_{\psi}$ reflects the distinct statistical behavior of the diagonal and the off-diagonal elements of $V^{N,N+p}$ in the $\psi_{\alpha}^{N}$ basis (within a perturbative approach, the SCHF single particle energy level position depends on both the diagonal and off-diagonal terms. In contrast the reorganization of SCHF wave functions depends only on the off-diagonal elements). 
%\begin{figure}[h] 
%\hspace{-1.5cm} \epsfig{figure=ss1.ps,width=9.cm}

%\vspace{-8.8cm} \hspace{+7.5cm} \epsfig{figure=ssca1.ps,width=9.cm}
%\caption{(a) Correlation function $\protect C_{\epsilon}(p)$ that
%measures the scrambling of single particle spectrum when $p$ particles
%are added. Size dependency is illustrated for $U=1.5,W=6$ and 
%$A=6 \times 6 \rightarrow 9\times 10$.
%Interaction and disorder strength dependencies look similar.
%(b) $C_{\epsilon}(p)$ as a
%function of the scaling variable $p/p_{\epsilon}$. For a given
%disorder $W$ and system area $A$, we plot all the data points corresponding
%to the entire $p$ range $p=\left\{ 1,2,..,10\right\} $ and interaction range 
%$U=\left\{ 0.5,1,..,3\right\} $.}
%\end{figure}

\section{Conclusion}
Based on a microscopic model of a disordered quantum dot, 
we have presented detailed numerical
results that show the importance of electron-electron interaction 
on the scrambling of the single particle properties obtained within a self
consistent Hartree-Fock method. As main result of this study, we have found
one-parameter scaling properties for the two quantities 
$C_{\psi }(p),C_{\epsilon}(p)$ which describe respectively the reorganization 
of the one particle wave functions and of the energy spectrum when $p$ 
particles are added to the system. The quantitative analysis 
of this one parameter scaling properties cannot be described by effective GRMP models considered until now \cite{scramblingtheo,grmp}. This suggests that 
new specific random matrix models are necessary 
to describe microscopic disordered models of Coulomb interacting particles 
within the Hartree-Fock approximation.

\end{document}